\documentclass{iopart}

\usepackage{amssymb}
\usepackage{amsthm}
\usepackage{graphicx}
\usepackage{subfig}
\usepackage{color}

\begin{document}

%\shorttitle{Generating functions for Vortices in BECs}
\title[Generating functions for Vortices in BECs]{Generating Functions, Polynomials and Vortices with Alternating Signs 
in Bose-Einstein Condensates}
\author{Anna M. Barry$^1$, F. Hajir$^2$, P.G. Kevrekidis$^{2}$}
\address{$^1$ Institute for Mathematics and its Applications, University
of Minnesota, Minneapolis, MN 55455}
\address{$^2$ Department of Mathematics and Statistics, University of Massachusetts,
Amherst MA 01003-4515, USA}

\begin{abstract}
In this work, we construct suitable generating functions for vortices
of alternating signs in the realm of Bose-Einstein condensates.
In addition to the vortex-vortex interaction included in earlier
fluid dynamics constructions of such functions, the vortices here
precess around the center of the trap. This results in the generating
functions of the vortices of positive charge and of negative charge
satisfying a modified, so-called, Tkachenko differential equation.
From that equation, we reconstruct collinear
few-vortex equilibria obtained in earlier work, as well as extend them
to larger numbers of vortices. Moreover, particular moment conditions
can be derived e.g. about the sum of the squared locations of the vortices
for arbitrary vortex numbers. Furthermore, the relevant differential
equation can be generalized appropriately in the two-dimensional
complex plane and allows the construction e.g. of polygonal vortex
ring and multi-ring configurations, as well as ones with rings surrounding
a vortex at the center that are again connected to earlier bibliography.
\end{abstract}

\section{Introduction}
\label{sec:intro}

The examination of vortex ``particles'' is a remarkable theme
of research in its own right both for smaller and even for larger
values of the number of vortices
$n$ that has stirred considerable interest in the fluid
dynamics engineering, as well as mathematical community. It is in
that light that it has been characterized as a ``classical mathematics
playground''~\cite{aref00}. One of the particularly enticing aspects
of this research activity has been the connection with the theory
of classical (such as the Hermite) and even more modern (so-called
Adler-Moser) polynomials. An excellent summary of this link
that interweaves applied mathematics with algebra and the theory of
polynomials is given in~\cite{aref11}; see also~\cite{arefextra,newton01}.

Recently, a new setting has emerged whereby vortices naturally
arise in experiments in quasi-two-dimensional systems. That involves
the dynamics of matter waves in ultra-cold atoms and more
specifically in the theme of Bose-Einstein 
condensates (BECs)~\cite{pethick,stringari,emergent}. There, the ability
to stir the system (imparting angular momentum), or to quench it
(spontaneously locking in vortex patterns), or to drag obstacles
through it (producing vorticity through the breakup of superfluidity)
enables the spontaneous production of one or more vortices, as has
now been summarized in numerous 
works~\cite{donnely,fetter1,fetter2,usmplb,tsubota}. In fact,
experiments have also enabled the production of vortices of higher
topological charge~\cite{middel16}, as well as the generation
of robust triangular vortex  lattices~\cite{middel13}.
In the early efforts along this direction, clusters of
few vortices were proposed theoretically~\cite{castindum}
and created experimentally~\cite{middel8}, however the emphasis
was chiefly on individual vortices and on large-scale vortex
lattices. Recent experimental work, however, has provided a renewed
focus on 
few-vortex clusters~\cite{bpa10,freilich10,bagn,Middelkamp2011,navar13}
through a variety of novel experimental techniques that have
offered a more controlled production of such clusters.

It is at the interface of these two fields that the present work
treads. In particular, the BEC realm has significant similarities
to the earlier fluid vortex setup in that BEC vortices interact in 
a way that can be accurately approximated by fluid  vortex
interactions~\footnote{This is progressively exact in higher number
of atom, more weakly trapped BECs; the inhomogeneous density due to the
trapping of BECs weakly screens this interaction, however we will not
be concerned with this effect here. An exposition of the screening
variation of the vortex interactions can be found in~\cite{busch} and
examples of how this can be captured (by effectively
altering the interaction term prefactor) without structurally 
altering the dynamical terms in the equations of motion can
be found e.g. in~\cite{Middelkamp2011,navar13}.}. However, our
setup has also a key additional ingredient affecting the vortex
motion, namely the motion of the vortices induced by the
(parabolic) magnetic trap. This external potential confining 
the atoms has been theoretically examined (see~\cite{fetter1}
for a relevant discussion reviewing the analysis) and experimentally
confirmed (see~\cite{freilich10} for a recent experimental corroboration)
to induce a precession of the vortices around the trap center. 
The associated precession frequency is nearly constant throughout,
roughly, half of the spatial extent of the BEC~\cite{navar13} and
for reasons of mathematical simplicity and tractability we will
consider it as constant in what follows~\footnote{However,
as we approach the rims of the BEC, this frequency starts to
rapidly increase as shown e.g. in~\cite{navar13}. A relevant discussion
in terms of accounting for this increase (and, more generally,
a radially dependent frequency) in future work is given in the
final section of the present work.}.

With the above described setup in mind (combining precession and 
inter-vortex interactions), we explore the stationary configurations
of vortices in BECs both in the case of co-rotating vortices~\cite{navar13}
(where only rigidly rotating configurations can be identified as steady
in a co-rotating frame; see below) and especially so in that of 
counter-rotating vortices. The former case reduces to a well-known example
from the fluid mechanical case, as analyzed in Section 2 (where it is
given as a preliminary calculation along the vein that we follow). However,
the counter-rotating case is more complex than its previously analyzed
fluid-mechanical sibling due to the presence of the precession term.
Hence, there we define two generating functions, one ($P$) with roots at
the positive vortex locations  and another ($Q$) 
with roots at the negative vortex
locations  and establish an ordinary differential equation satisfied
by a combination of the two for vortices along a real line (section 3;
an alternative derivation using a rational distribution function is
given in section 4)
and within the complex plane (section 6). In section 5, we then use it
to produce specific examples of stationary collinear
vortex configurations, both
ones that are more standard and have been examined before (including in
experiments~\cite{bagn,Middelkamp2011}) and ones that have not.
In the complex plane case of section 6, we identify polygonal vortex
ring and multi-ring configurations (as well as ones with such
rings surrounding a centrally located vortex)
and also connect them with earlier numerical
observations. Finally, in section 7, we summarize our findings and
present a number of relevant conclusions, as well as a series of
open problems for future work.

\section{Setup and a Preliminary Calculation: Collinear Co-rotating
Vortices}
\label{sec: setup}

In the present study, we briefly touch upon this theme, connecting
it to the considerably more complex setting considered herein whereby
the vortices feature precession in addition to their pairwise
interactions. Perhaps the most elementary yet particularly elegant
calculation associated with both settings can be given in the case
of $n$ vortices of the same charge, when considering solidly rotating
solutions thereof. In that case, the
equation of motion reads:
\begin{eqnarray}
\Omega x_j = \sum_{1\leq i\neq j \leq n} \frac{1}{x_j - x_i} \qquad 1\leq j \leq n.
\label{rot}
\end{eqnarray}
In this setting the precession frequency $\Omega$ 
effectively renormalizes the vortex positions (which scale
as $\propto 1/\sqrt{\Omega}$), hence we will set it to unity
in what follows. Furthermore, the $\Omega$ term is
the one accounting for the rotational frequency of the
entire configuration and if an in-trap precession term
exists, it can be directly absorbed into this ``lumped''
precessional term.

The remarkable idea that apparently dates back to the
work of Stieljes~\cite{s1885} (subsequently developed
by Marden~\cite{marden}, and Szeg{\"o}~\cite{szego}, among others)
was to consider a, so-called, generating function in the form of:
\begin{eqnarray}
P(x)=\prod_{i=1}^{n} (x-x_i) = (x-x_1) \dots (x-x_n)
\label{gener}
\end{eqnarray}
Then, straightforward differentiation leads to the identities:
\begin{eqnarray}
\hspace{-10mm} P'(x)=P(x) \sum_{i=1}^n \frac{1}{x-x_i}, \quad
P''(x)=P(x) \sum_{i=1}^{n} \sum_{j=1,j\neq i}^n \frac{1}{(x-x_i) (x-x_j)}
\label{gener1}
\end{eqnarray}
A key observation then is that the product involved in the second
derivative can be simply written as:
$\frac{1}{(x-x_i)(x-x_j)} = [\frac{1}{x-x_i}-\frac{1}{x-x_j}] \frac{1}{x_i-x_j}$.
But then the sum over $j$ in the  term $\frac{1}{(x-x_i) (x_i-x_j)}$
can be performed because
it is equal to $x_i/(x-x_i)$ according to Eq.~(\ref{rot}). The second bracket
is handled in the same manner (the summation over $i$ is performed)
and exchanging indices we get $P''(x)=2 P(x) \sum \frac{x_i}{x-x_i}$.
Of course this is true {\it only} provided that the points $x_i$
are the equilibrium vortex configuration locations. Then, rewriting
$x_i=-(x-x_i) + x$, we obtain a summation of (-) unity (yielding a factor
of $- n$ as multiplying $2 P(x)$), as well as a factor of $2 x P'(x)$,
from the identity of Eq.~(\ref{gener1}) for $P'(x)$. Combining all
the above, the differential equation 
\begin{eqnarray}
P''(x)=-2 n P(x) + 2 x P'(x)
\label{hermi}
\end{eqnarray}
arises for the generating function which is {\it exactly} the differential
equation satisfied by the n-th order Hermite polynomial. In light of
the above analysis, this implies that the roots of a rotating equilibrium
configuration involving $n$ vortices will lie exactly at the root locations
of the $n$-th order Hermite polynomial. 
 
\section{Generating Function ODE for
Collinear Counter-rotating BEC Vortices}
\label{sec: Generating}

We now turn to a variant of this analysis that is relevant to
our setting involving vortices also featuring precession.
Perhaps the most appropriate context for illustrating
relevant ideas is the case of equilibria of
precessing vortices involving opposite charges~\cite{barry13}.
In this case, we will start by assuming that there are $n_+$
vortices of one charge and $n_-$ vortices of the other charge.

Without loss of generality, we record the position of
negative charges with an odd index and the position of 
the positive charges with an even index, so that $x_j$
has charge $(-1)^j$ for all $j$.
Then, the steady state equations for the ``odd'' charges 
 reads:
\begin{eqnarray}
x_{2 k+1} = \sum_{l=1}^{n_+} \frac{1}{x_{2k+1} - x_{2l}}
- \sum_{l=1,l \neq k}^{n_-} \frac{1}{x_{2k+1} - x_{2 l+1}}.
\label{eqvor1}
\end{eqnarray}
Similarly for the ``even'' positive charges, we will have
\begin{eqnarray}
x_{2 k} = - \sum_{l=1,l \neq k}^{n_+} \frac{1}{x_{2k} - x_{2l}}
+ \sum_{l=1}^{n_-} \frac{1}{x_{2k} - x_{2 l+1}}.
\label{eqvor2}
\end{eqnarray}
Notice that here these are genuine equilibrium (i.e., stationary)
configurations rather than rotating ones, as the opposite 
actions of precession and interaction can ``balance'' each other out.

Then, we can define $P(x)=\prod_{l=1}^{n_+} (x-x_{2l})$
and $Q(x)=\prod_{l=1}^{n_-} (x-x_{2 l+1})$, which satisfy
similar identities to Eq.~(\ref{gener1}) [since these
do not involve in any way the equations of motion].
Using then the same partial fraction decomposition as above
and the equations of motion~(\ref{eqvor1}) and~(\ref{eqvor2}),
we obtain:
\begin{eqnarray}
\hspace{-10mm} P''(x) &=& 
2 P(x) \left[ \sum_{l=1}^{n_+}\frac{-x_{2l}}{x-x_{2 l}} + \sum_{l=1}^{n_+} 
\sum_{k=1}^{n_-} \frac{1}{(x-x_{2 l}) (x_{2l}-x_{2k+1})} \right]
\label{eqvor3}
\\
\hspace{-10mm} Q''(x) &=& 
2 Q(x) \left[ \sum_{k=1}^{n_-}\frac{-x_{2k+1}}{x-x_{2 k+1}} + \sum_{l=1}^{n_+} 
\sum_{k=1}^{n_-} \frac{1}{(x-x_{2 k+1}) (x_{2k+1}-x_{2l})} \right].
\label{eqvor4}
\end{eqnarray}
Multiplying $Q \times$ (\ref{eqvor3}) $+ P \times$ (\ref{eqvor4}),
and once again reshuffling according to our well-known by now 
partial fraction identity,
we retrieve a differential equation combining the generating 
functions $P$ and $Q$ in the form:
\begin{eqnarray}
P Q'' + Q P''=2 P' Q' + 2 (n_+ + n_-) P Q - 2 x (P Q)'.
\label{eqvor5}
\end{eqnarray}
Remarkably, such equations in the simplest setting of vortices
purely interacting (but not precessing) had been developed previously stemming
already from the work of Tkachenko in 1964; see the relevant discussion
in~\cite{aref11}.
However, the more standard case of purely interacting vortices
provides a much simpler form of Eq.~(\ref{eqvor5}) as
$P'' Q + P Q''=2 P' Q'$. An elegant derivation of this latter
form also appears in the work of~\cite{campbell}.

We should additionally note here that in the above calculation 
we labeled the vortices
as even and odd implicitly 
suggesting that the number of the former and of the latter
is either equal or differs by one. However, it is straightforward to 
observe that nowhere in the above derivation is this a crucial assumption
and in principle the methodology applies for all vortex numbers, hence 
so does Eq.~(\ref{eqvor5}). Nevertheless, it is important to point out 
that for vortices of equal charge magnitude (as considered here), 
we have been unable to find such solutions in our special case
examples considered below (e.g. with $3$ vortices of one charge and
$1$ of the other, or $4$ of one and $1$ of the other etc.). Hence,
we will not consider such configurations further here and instead 
will turn, in section 6, to special case examples of either equal or differing by one vortex
numbers of opposite charges. However, there will also be in that 
section a notable example of $n$ vortices of unit charge and one
vortex of a different opposite (i.e., not equal and opposite) charge.

\section{An Alternative Rational Function Derivation}
\label{sec: Rational}

Before turning to examples that illustrate the findings of the previous section, 
it might be useful for further investigations to record an alternative
derivation of the key equation of section 3 which uses a \emph{rational}
function as opposed to a polynomial.

Suppose $s_1,\ldots, s_n$ are non-zero integers and $x_1,\ldots, x_n$ are distinct real numbers satisfying 
for each $j\in \{1,2,\ldots, n\}$,
\begin{equation}
\label{fundamental-equality}
\sum_{1\leq i \neq j \leq n}\frac{s_i}{x_i-x_j} = \epsilon s_j x_j, 
\end{equation}
where we consider $\epsilon= \pm 1$.
In this formulation, the $\epsilon=+1$ case is relevant for the realm of
co-rotating vortices as per Eq.~(\ref{rot}), 
when all $s_i=1$.
 The $\epsilon=-1$ sign case
is the proper one for the case of stationary (non-rotating)
configurations of counter-rotating vortices, as 
per Eqs.~(\ref{eqvor1})-(\ref{eqvor2}). More specifically,
if we set $s_i=(-1)^i$ for all $i$, we are exactly in the situation of (\ref{eqvor1}) and (\ref{eqvor2}) of section \ref{sec: Generating}.
Consider the rational function $$R(x)=\prod_{i=1}^n (x-x_i)^{s_i}.$$  By logarithmic differentiation, we have
$$
R'(x) = R(x) \sum_{i=1}^n \frac{s_i}{x-x_i}.
$$
Differentiating the above equation and dividing by $R(x)$, we find
\begin{eqnarray*}
\frac{R''(x)}{R(x)} &=& \frac{R'(x)}{R(x)} \sum_{i=1}^n \frac{s_i}{x-x_i} - \sum_{i=1}^n \frac{s_i}{(x-x_i)^2} \\
&=& \sum_{1\leq i \neq j \leq n} \frac{s_i s_j}{(x-x_i)(x-x_j)} + \sum_{i=1}^n \frac{s_i^2 - s_i}{(x-x_i)^2}.
\end{eqnarray*}

Applying the standard partial fraction decomposition
%$$
%\frac{1}{(x-x_i)(x-x_j)} = \left( \frac{1}{x-x_i} - \frac{1}{x-x_j} \right) \frac{1}{x_i - x_j}
%$$
to the first sum on the right hand side above, and using (\ref{fundamental-equality}), we find after a small calculation that
\begin{eqnarray}
\nonumber
\frac{R''(x)}{R(x)} &=& \epsilon 2 \sum_{i=1}^n \frac{s_i^2 x_i}{x - x_i} 
- \sum_{i=1}^n \frac{s_i^2 - s_i}{(x-x_i)^2}.
\\
\label{2nd}
&=& \epsilon \left(-2n + 2x \sum_{i=1}^n \frac{s_i^2}{x-x_i}\right)
- \sum_{i=1}^n \frac{s_i^2 - s_i}{(x-x_i)^2}.
\end{eqnarray}
For the $\epsilon=1$ sign case and $s_i=1$ for all $i$, when equation (\ref{fundamental-equality}) reduces simply to (\ref{rot}), then $R(x)$ is a polynomial with $n$ distinct roots, and we recover $R''(x)/R(x) = -2n + 2x \sum_{i=1}^n \frac{1}{x-x_i}$, in agreement with our derivation of section 2.

Hereafter, we focus on the $\epsilon=-1$ sign case, per the configurations
of section 3, involving counter-rotating vortices.
Here $R$ is not a polynomial, and we proceed to separate it into its numerator and denominator, writing $R(x)=P(x)/Q(x)$ with polynomials
$$
P(x) = \prod_{1\leq i \leq n: s_i>0} (x-x_i)^{s_i}, \qquad
Q(x) =\prod_{1\leq i \leq n: s_i<0} (x-x_i)^{-s_i}.
$$
Note that the convention of the previous section fits into this setup simply with the assignment $s_i=(-1)^i$ for all $i$.
If we let $Y(x)=P(x)Q(x)$, then (\ref{2nd}) above can be written in a 
compact form, namely
$$
\frac{R''(x)}{R(x)} = 2n - 2x \frac{Y'}{Y} + 2 \left(\frac{Q'}{Q}\right)'.
$$
We expand the right hand side and multiply both sides by $PQ$ to find
%$$
%Q^2R'' = 2nPQ + 2x (PQ)' + 2Q''P -2RQ'^2.
%$$
(using also that $Y'' = P''Q+2P'Q'+PQ''$) finally that
%under (\ref{fundamental-equality}),
$$
Y''+2xY'-2nY=4P'Q'.
$$
This is the same modified Tkachenko equation as obtained in section 3.

\section{Special Case Examples}
\label{sec: Special}

Let us now seek special case example solutions of Eq.~(\ref{eqvor5})
to explore the power of the generating function formalism in comparison
to the brute force calculations, based on the equations of motion
in~\cite{barry13}. The simplest case to examine is that of
monomials with $P(x)=(x-a)$ and $Q(x)=(x+a)$, as per the vortex
dipole solution examined previously; see also the recent
experiments of~\cite{freilich10,Middelkamp2011}. In this case, we have
$Q(x)=- P (-x)$; in fact, more generally for even vortex number
with $n_+=n_-=n$, we have that $Q(x)=(-1)^n P(-x)$, based on the
symmetries discussed in~\cite{barry13}. A direct substitution
of $P$ and $Q$ into Eq.~(\ref{eqvor5}) yields immediately $a= \pm 1/\sqrt{2}$.
This, in turn, implies that $P(x) Q(x)=(x^2-a^2)=(x^2-1/2) \equiv H_2(x)$
i.e., it appears that the Hermite polynomials resurface, albeit
now in the form of $P(x) Q(x)$. Unfortunately, this is only
a fortuitous coincidence, as in this case of monomials $P''=Q''=0$
and hence one can proceed to show that 
\begin{eqnarray}
(P Q)''=2 P' Q'=2 x (P Q)' +4 P Q.
\label{h2case}
\end{eqnarray}
This, in turn, establishes that $P Q = H_2(x)$ according to the equation
for the relevant polynomials (cf. Eq.~(\ref{hermi})). 

In the case of odd $n_+ + n_-$, the above symmetry of the
generating functions is broken and we have e.g. for $n_+=1$
and $n_-=2$ that $P(x)=x$ and $Q(x)=x^2-a^2$. Once again, a very
straighforward direct substitution yields $a= \pm 1/\sqrt{2}$.

On the other hand, for $n_+=n_-=2$, we have a choice of
$P(x)=(x-a) (x-b)$ and $Q(x)=(x+a) (x+b)$. In this case,
direct substitution of the $P$ and $Q$ in Eq.~(\ref{eqvor5})
yields the algebraic conditions:
\begin{eqnarray}
a^2 + b^2 =1, \quad a b=\frac{1 \pm \sqrt{2}}{2}
\label{eqvor6}
\end{eqnarray}
which is in line with the corresponding finding of~\cite{barry13}.

In the case of $n_+=3$ and $n_-=2$, the generating functions
assume the following form $P(x)=x (x^2-a^2)$ and $Q(x)=x^2 -b^2$.
Here, the direct substitution yields anew a system for $a, b$
according to:
\begin{eqnarray}
a^2 + b^2=1, \quad 2 a^2 - 6 b^2 +8 a^2 b^2 =0.
\label{eqvor7}
\end{eqnarray}
This, in turn, results in $a^2=\sqrt{3}/3$ and $b^2=(2-\sqrt{3})/2$, which
we again find to be in agreement with~\cite{barry13}.

The last two cases that we wish to consider following this method
are $n_+=3$ and $n_-=3$, as well as $n_+=3$ and $n_-=4$. In the former
case, $P(x)=(x-a) (x-b) (x-c)$ and $Q(x)=-P(-x)$. This results in
three algebraic equations of the following form:
\begin{eqnarray}
a^2 + b^2 + c^2 &=& \frac{3}{2}
\label{eqvor8}
\\
12 a^2 b^2 c^2 -2 (a^2 b^2 + a^2 c^2 + b^2 c^2)- 8a b c (a+b+c) &=& 0
\label{eqvor9}
\\
4 (a+b+c)^2 - 8 (a^2 b^2 + b^2 c^2 + a^2 c^2) &=& 0
\label{eqvor10}
\end{eqnarray}
The solution of the algebraic system of Eqs.~(\ref{eqvor8})-(\ref{eqvor10})
is $a=0.1623$, $b=-0.4765$ and $c=1.1165$, which indeed when scaled using
$\Omega=1/2$ as in~\cite{barry13}, yields values identical to the
ones reported therein.

 Finally, the case of $n_+=3$ and $n_-=4$ yields $P(x)=x (x^2-a^2)$
and
$Q(x)=(x^2-b^2) (x^2-c^2)$. Here, the algebraic equations yield:
\begin{eqnarray}
a^2 + b^2 + c^2 &=& \frac{3}{2}
\label{eqvor11}
\\
4 (b^2+c^2-a^2) -8 (a^2 b^2 + a^2 c^2 + b^2 c^2) &=& 0
%12 a^2 b^2 c^2 -2 (a^2 b^2 + a^2 c^2 + b^2 c^2)- 8a b c (a+b+c) &=& 0
\label{eqvor12}
\\
12 a^2 b^2 c^2 -2 (a^2 b^2 + a^2 c^2 -3 b^2 c^2) &=& 0.
%4 (a+b+c)^2 - 8 (a^2 b^2 + b^2 c^2 + a^2 c^2) &=& 0
\label{eqvor113}
\end{eqnarray}
These yield $a=0.5546$, $b=0.2594$ and $1.0607$ (again potentially
scalable by $\sqrt{1/\Omega}$).

For reasons of completeness, it is relevant to add here
a discussion about the stability and observability
of such configurations. Among the above aligned vortex
configurations, the only one that is dynamically robust
and observed experimentally as such is the vortex dipole with 
$n_+=n_-=1$; see e.g.~\cite{freilich10,Middelkamp2011} and references therein.
Dipole dynamics (although not at steady state) have been also
experimentally observed in~\cite{bpa10}. All higher aligned 
configurations, even though stationary, are, in fact, dynamically unstable,
as has been illustrated through detailed numerical bifurcation
computations in~\cite{middel10}. This is a direct byproduct of
their emergence through a cascade of super-critical
pitchfork bifurcations from the destabilization of
the two-dimensional (planar) analog of a dark soliton.
The first such instability (that endows the dark soliton
state with one unstable mode) produces as a ``daughter branch''
the stable vortex dipole. At the next instability, the
vortex tripole emerges ($n_+=1$ and $n_-=2$ or equivalently $n_+=2$ and
$n_-=1$), although the already present
instability of the dark soliton endows it already with 
an unstable eigenmode, while the dark soliton itself now
possesses two unstable eigenmodes past the bifurcation point.
Then the cascade continues by producing the aligned quadrupole
with $n_+=2$ and $n_-=2$ which will now inherit the dark
soliton's two unstable modes, while the dark soliton gets
endowed with a third one and so on and so forth. 
However, it should be pointed out that despite the instability
of all higher states with $\max(n_+,n_-)>1$, these states are
still potentially observable as long-lived metastable states.
An example of this type is the vortex tripole which has been
observed as a long-lived state in the experiments of~\cite{bagn}.
Additionally, techniques such as the ones of~\cite{bpa10}, of
laser beams creating potential barriers can be utilized in
order to create (and trap) 
different numbers of vortices within the condensates at will~\cite{bpa14}.
Along that experimental vein, one can straightforwardly envision
producing stationary configurations of higher $n$ and
observing their dynamics (and unstable evolution). As
an additional possibility of potential experimental
interest, it should be mentioned that such aligned
vortex configurations have also been identified in
anisotropic settings~\cite{janstock}. There, depending
on the strength of the anisotropy in the
transverse (to the line bearing the vortices) direction,
it has been demonstrated to be possible to stabilize
configurations with arbitrarily many aligned
counter-rotating vortices.

It is interesting to add here that observation of the above 
patterns of algebraic equations suggests that it may
be possible to derive general algebraic conditions that
the roots of the polynomials may satisfy irrespectively
of $n_+$ and $n_-$. For instance, we can see that both
for $n_+=2=n_-$ and for $n_+=3=n_-+1$, it is true that
$\sum_{i=1}^{n_++n_-} x_i^2=1$, while the same sum becomes
$3/2$ for $n_+=3=n_-$ and for $n_+=4=n_-+1$. Motivated
by this finding, we have more generally examined
the resulting powers of the polynomials arising
in both the left and the right hand side of Eq.~(\ref{eqvor5}).
As a result, we have found that at power $x^{n_++n_--2}$,
the prefactors of the two sides are respectively
$n_+ (n_+-1) + n_- (n_--1)$ and
$-4 \sum_{i=1}^{n_++ n_-} x_i^2 +2 n_+ n_-$, leading eventually
to the identity:
\begin{eqnarray}
\sum_{i=1}^{n_++ n_-} x_i^2 =\frac{1}{4} \left[2 n_+ n_- - n_+ (n_+-1)
- n_- (n_--1) \right].
\label{eqvor14}
\end{eqnarray}

Such identities (verified by all of our above considered cases)
have also been considered in the absence of precession
in~\cite{aref11} (see e.g. Eq.~(10) therein), although they are obtained
at the level of the dynamical equations for the vortices rather than
those of the generating function ODEs. 

\section{Generalizing the Tkachenko Equation in the Complex Plane}
\label{sec: complex}

Among the
numerous aspects that are worthy of additional investigation
in the realm of oppositely charged vortices in the presence
of precession, arguably, one of the most important 
concerns the generalization of the above considerations to states that
are not collinear.
Hence, we now turn our attention to generalizing the above considerations
of our modified Tkachenko equation to the two-dimensional
complex plane. 

In this case, if we take $n_+$ vortices $z_1,..,z_{n_+}$ with charge $+1$ and $n_-$ vortices $\zeta_1,...,\zeta_{n_-}$ with charge $-1$ the relevant stationary equations will be of the form:
\begin{eqnarray}
z_j &= -\sum_{k=1,k \neq j}^{n_+} \frac{z_j-z_k}{|z_j-z_k|^2}+\sum_{k=1}^{n_-} \frac{z_j-\zeta_k}{|z_j-\zeta_k|^2}
\\
\zeta_j&=-\sum_{k=1,k \neq j}^{n_-} \frac{\zeta_j-\zeta_k}{|\zeta_j-\zeta_k|^2}+\sum_{k=1}^{n_+} \frac{\zeta_j-z_k}{|\zeta_j-z_k|^2}.
\label{eqnew1}
\end{eqnarray}
We again define the $P$ and $Q$ generating functions in the same way
as before as $P(z)=\prod_{i=1}^{n_+} (z-z_i)$ and $Q(z)=\prod_{j=1}^{n_-}
(z-\zeta_j)$. While Eqs.~(\ref{gener1}) remain the same in this case, the
main change arises at the level of Eqs.~(\ref{eqvor3})-(\ref{eqvor4}).
Here, the identities become
\begin{eqnarray}
P''(z) &=& P(z) \sum_{i=1}^{n_+} \frac{2}{z-z_i} \left[ -\bar{z}_i + \sum_{k=1}^{n_-}
\frac{1}{z_i-\zeta_k} \right]
\label{eqnew2}
\\
Q''(z) &=& Q(z) \sum_{j=1}^{n_-} \frac{2}{z-\zeta_j} \left[ -\bar{\zeta}_j + \sum_{k=1}^{n_+}
\frac{1}{\zeta_j-z_k} \right]
\label{eqnew3}
\end{eqnarray}
In a way paralleling the derivation of Eq.~(\ref{eqvor5}), we 
form $P''(z) Q(z) + Q''(z) P(z)$. In that context, the second term
in the brackets within Eqs.~(\ref{eqnew2})-(\ref{eqnew3}) will again
yield $2 P'(z) Q'(z)$, but the first term is more complicated.
In particular, $\bar{z}_j=x_j - i y_j = z_j - 2 i y_j$ and $\bar{\zeta}_j=\zeta_j-2i\upsilon_j$. The first
term (involving $z_j$ and $\zeta_j$) is now entirely analogous to the real case and will
provide the terms $2 (n_+ + n_-) P(z) Q(z) - 2 z (P(z) Q(z))'$.
However, the last term provides an additional term $4 i P(z) Q(z) 
\left(\sum_{j=1}^{n_+} \frac{y_j}{z-z_j}+\sum_{j=1}^{n_-} \frac{\upsilon_j}{z-\zeta_j}\right)$ (cf. the term of the right hand
side of Eq.~(12) of~\cite{aref11}). The resulting final form of
the modified Tkachenko equation is:
\begin{eqnarray}
\hspace{-10mm} P'' Q + P Q'' &=& 2 P' Q' + 2 (n_+ + n_-) P Q -2 z (P Q)' 
\nonumber
\\
&+& 4 i P Q\left(\sum_{j=1}^{n_+} \frac{y_j}{z-z_j}+\sum_{j=1}^{n_-} \frac{\upsilon_j}{z-\zeta_j}\right)
\label{eqnew5}
\end{eqnarray}

Generally, Eq.~(\ref{eqnew5}) is far more difficult to solve than
Eq.~(\ref{eqvor5}). However, relevant results can still be extracted
from it in some special
case examples. Perhaps a prototypical one is that of two interlaced
polygonal vortex rings (with the same radius) 
consisting of oppositely charged vortices. Non-stationary variants
of such configurations were observed to spontaneously emerge through
the instability of ring dark solitons in~\cite{RDS}. Stationary variants
of such states were identified in~\cite{toddricardo} and a more
systematic bifurcation analysis thereof connecting them to the
instabilities of the ring dark soliton was given in~\cite{middel11b}.
An analysis of the existence and stability of these configurations
for arbitrary total number of vortices $n_+ + n_-$ was given in~\cite{barry13}.
It was found there that the only (non-collinear) such configuration 
which is stable at the ``particle level'' is the so-called
vortex quadrupole~\cite{quadrupole}; see also~\cite{middel10}.
On the other hand, configurations such as the vortex hexagon,
octagon, decagon, etc. are always unstable, arising through a
cascade of progressive bifurcations from the ring dark soliton.
Interestingly, this is an entirely
analogous process to the cascade of instabilities
imparted on the collinear vortex states by the rectilinear
dark soliton.
This imparts additional unstable modes to the waveform as the
number of vortex increases i.e., the hexagon has always at least
one unstable mode, the octagon at least two, the decagon at least
three and so on. Moreover, the polygonal ring 
states are {\it absent} for the case where $n_+ + n_-$ is odd~\cite{barry13}.
Now, we turn to the identification of such 
states through the Tkachenko equation approach of Eq.~(\ref{eqnew5}).

We assume that $n_+=n_-\equiv n$ and then the two polygonal rings,
possessing the same radius can be represented in the complex
plane by $P(z)=z^{n}-R^n$ and $Q(z)=z^n-R^n e^{i n \phi}$. In the
second expression the roots (i.e., vortex locations) are assumed
to be (uniformly) shifted by an angle $\phi$. This achieves the 
interlaced vortex ring configuration of interest. Direct substitution
of the $P$ and $Q$ in Eq.~(\ref{eqnew5}) yields numerous cancelations
and only two conditions, due to the vanishing of the polynomial
prefactors of $z^{2 n -2}$ and of $z^{n-2}$
[it should be added as an aside that properties of the complex roots
of unity and geometric series are used in order to evaluate the last
term in Eq.~(\ref{eqnew5}); cf. Eq.~(13) of~\cite{aref11}]. 
These two conditions are:
\begin{eqnarray}
R^2 &=& \frac{1}{2}
\label{eqnew6}
\\
e^{i n \phi} &=& -1
\label{eqnew7}
\end{eqnarray}
The former condition represents an algebraic constraint either on the
radius of the vortex ring or, equivalently (for arbitrary vortex ring
radius) on the precession frequency (cf. with~\cite{barry13}). The latter
condition, however, is arguably even more important as it suggests 
that the two rings are displaced with respect to each other by
an angle of $\phi=\pi/n$. This is consonant with the numerical observations
(also of the corresponding PDE model), as e.g. in the case of the vortex
quadrupole~\cite{quadrupole,middel10} of $n=2$, the rotation is by
$\pi/2$, in that of the hexagon~\cite{toddricardo,middel11b} 
of $n=3$, it is by $\pi/3$ and so on. Moreover, since
the vortices of a single charge are rotated with respect
to their neighbors of the same sign by $2 \pi/n$ (due to the
structure of the complex roots of unity), this suggests
that in the interlaced polygonal configuration, (as expected by symmetry)
the vortices are positioned in an alternating charge fashion at equal
angles of $\pi/n$.

Beyond the single vortex polygon one can consider two or more nested polygons.  Two rings, each possessing $4$ vortices, were observed numerically in the PDE model as bifurcating from the X-shaped dark soliton cross ~\cite{middel11b}.  Like in the case of a single polygon, we find that there are no fixed equilibria consisting of nested polygons with odd numbers of vortices.  With this as motivation we investigate two polygons (each interlaced as above) having an equal, even number of vortices.  We consider  Eq.~(\ref{eqnew5}) with
\begin{eqnarray}
P(z)&=(z^n-R_1^n)\left(z^n-R_2^n e^{i n\phi}\right)
\\
Q(z)&=(z^n-R_2^n)\left(z^n-R_1^n e^{i n\phi}\right)
\end{eqnarray}
and obtain a polynomial in $z$ which, after factoring out $z^{n-2}$, has only four (potentially) nonzero terms:
\begin{eqnarray}
\hspace{-20mm} &4 n (-1 + R_1^2 + R_2^2) z^{4 n}
\nonumber
\\
\hspace{-20mm} &-n(1 + e^{i n \phi}) (R_2^n (-3 + n + 4 R_1^2 + 2 R_2^2) + 
   R_1^n (-3 + n + 2 R_1^2 + 4 R_2^2)) z^{3 n}
\nonumber
\\
\hspace{-20mm} &+ 2 ne^{i n \phi} ((-1 + 2 R_1^2) R_2^{2 n} + R_1^{2 n} (-1 + 2 R_2^2) 
\nonumber
\\
\hspace{-20mm} &+ 
   2 R_1^n R_2^n (-1 + 2 n + R_1^2 + R_2^2 + (-1 + R_1^2 + R_2^2) \cos(n \phi))z^{2n}
\nonumber
\\
\hspace{-20mm} &-e^{ i n \phi} (1 + e^{i n \phi}) n R_1^n R_2^n ((-1 + n + 2 R_1^2) R_2^n + 
   R_1^n (-1 + n + 2 R_2^2)) z^n.
\end{eqnarray}
Setting the first expression equal to zero we find the relationship between the two radii $R_1^2+R_2^2=1$.  Using this relationship and noting that $n\geq 2$ forces us to choose $\phi=\pi/n$ from the second term.  This choice also renders the fourth term identically zero, and we now have two equations which can be used to determine admissible radii for the rings, depending on the
particular value of $n$:
\begin{eqnarray}
1&=R_1^2+R_2^2\\
0&=4 n R_1^n R_2^n + (-1 + 2 R_1^2) R_2^{2 n} + R_1^{2 n} (-1 + 2 R_2^2).
\end{eqnarray}

Interestingly, we are able to modify the forms of $P$ and $Q$ slightly to deduce the existence of another family of equilibria made up of two interlaced $N$-gons with different radii and a vortex at the center of the configuration.  A configuration from this family can be found in ~\cite{middel11b}, Figure 6(b), upper left.  Without loss of generality, we choose the vortex at the center to have charge $+1$ and set
\begin{eqnarray}
P(z)&=z(z^n-R_1^n)
\\
Q(z)&=\left(z^n-R_2^n e^{i n\phi}\right).
\end{eqnarray}
Now, Eq.~(\ref{eqnew5}) is a polynomial with two nonzero coefficients corresponding to $z^n$ and $z^{n-1}$.      The coefficients are zero provided
\begin{eqnarray}
1&=R_1^2+R_2^2\\
0&=-e^{i n \phi} (1 + n + 2 R_1^2) R_2^n - R_1^n (-3 + n + 2 R_2^2).
\end{eqnarray}
By dividing the second equation into real and imaginary parts, we again find $\phi=\frac{\pi}{n}$.  For a given $n$ the remaining equations can again
be used to determine $R_1$ and $R_2$. Both the previous and this
configuration are generalizations and analytical characterizations 
of specific example solutions identified
at the level of the partial differential equation 
(the so-called Gross-Pitaevskii model) describing the mean-field atomic
BEC system~\cite{middel11b} for particular $n$'s ($n=2$ for
the former, and $n=4$ for the latter case).

There exists one more case in which we were also
able to generalize the calculation in the complex plane. 
This pertains to the situation where a single vortex (to be eventually
placed at the origin) is surrounded by vortices of (what we will
see has to be by necessity, if the configuration is to be stationary)
opposite charge. It is noteworthy that this case can also be 
captured by Eq.~(\ref{eqnew5}) upon suitable modification (to encompass
the general charge at the origin and the choice of $Q(z)=z$), however,
we will develop the relevant case directly from the stationary equation
for the system of a single vortex of charge $S_0$ located at $z_0$,
surrounded by $n$ vortices of charge $S$ located on a ring of radius $R$.
In that case, assuming a unit frequency of precession, we have that:
\begin{eqnarray}
S_0 z_0 &=& - \sum_{j=1}^n \frac{S}{\bar{z}_0-\bar{z}_j}
\label{extr1}
\\
S z_k &=&- \sum_{j=1, j\neq k} \frac{S}{\bar{z}_k-\bar{z}_j} 
-\frac{S_0}{\bar{z}_k-\bar{z}_0}
\label{extr2}
\end{eqnarray}
We define as previously $P(z)=\prod_{j=1}^n (z-z_j)$, having
as before that $P''(z)=2 P(z) \sum_{i=1}^n \sum_{j=1, j\neq i}
\frac{1}{z-z_i} \frac{1}{z_i-z_j}$. Performing the step
that is by now fairly familiar, the summation
over $j$ can be substituted (from Eq.~(\ref{extr2})) yielding:
\begin{eqnarray}
P''(z)=2 P(z) \sum_{i=1}^n \sum_{j=1, j\neq i}
\frac{1}{z-z_i} \left[-\frac{S_0}{S} \frac{1}{z_i-z_0}
- \bar{z}_i \right]
\label{extr3}
\end{eqnarray}
The last term will yield familiar contributions in the equation
leading to terms (cf. Eq.~(\ref{eqnew5})): $2 n P(z) -2 z P'(z)
+ 4 i P(z) \sum_n y_i/(z-z_i)$. On the other hand, the contribution 
of the first term is less familiar and amounts to
$-2 P(z) (S_0/S) \sum_{i=1}^n (z-z_i)^{-1} (z_i-z_0)^{-1}$. Using the
fact that $P(z)=z^n-R^n$ amounting to a ring of $n$ vortices of
charge $S$ surrounding the central one of charge $S_0$, all terms
other than then one proportional to $S_0/S$ in Eq.~(\ref{extr3})
can be computed. The term $\propto S_0/S$ can be computed, however,
as well upon the additional assumption that $z_0=0$, which is 
also meaningful from the point of view of the configuration's
symmetry. In that case, we have that:
\begin{eqnarray}
\hspace{-10mm} -2 P(z) \frac{S_0}{S} \sum_{i=1}^n \frac{1}{z-z_i} \frac{1}{z_i-z_0}
= -2 P(z) \frac{S_0}{S} \frac{n z^{n-2}}{z^n-R^n} =-2 n \frac{S_0}{S} n z^{n-2}
\label{extr4}
\end{eqnarray}
Using this identity and the form of $P(z)$, direct substitution 
in Eq.~(\ref{extr3}) yields [in addition to numerous cancellations]
the formula:
\begin{eqnarray}
R^2= -\frac{S_0}{S} + \frac{1-n}{2}
\label{extr5}
\end{eqnarray}
From this, we naturally infer that $S_0$ and $S$ must be oppositely
charged i.e. $S_0 S < 0$ for such a stationary configuration to exist.
This is also intuitively clear, as otherwise (if all charges were
of the same sign), the configuration would tend to rigidly rotate
rather than be at equilibrium. In fact, since the contribution of
$(1-n)/2$ is negative, what Eq.~(\ref{extr5}) suggests is that not
only should $S_0$ be of opposite sign than $S$, but also of sufficiently
high charge, so as to effectively counter the rotational tendency
of the charges $S$. I.e., the relevant equilibrium condition, in 
addition to the charge sign inequality above, yields $|S_0/S|> (n-1)/2$.
It is interesting to note here that a special case of this 
configuration with $n=4$ was encountered previously in~\cite{barry13}
(see Eq.~(13) therein), as well as numerically analyzed
(again for $n=4$) in~\cite{middel11b}, where its instability was
corroborated. 

While in the above analysis, for reasons of analytical tractability and
computational convenience, we have restricted our considerations to the
case of symmetric polygonal configurations, there exist numerous
intriguing examples of additional stationary configurations.
More specifically, as can be seen in~\cite{aref0} for the case
where the precession term is absent, 
there are numerous configurations that are genuinely asymmetric
(cf.~\cite{aref0}, Fig. 4 in~\cite{aref11} etc.). It would be of
particular interest (although it would most likely be a  numerical
task) to examine the possibility of such configurations in the present
context.

\section{Conclusions and Future Challenges}
\label{sec: Conclusion}

In the present work, we extended the considerations of~\cite{aref11}
to the realm of Bose-Einstein condensates and the cases of vortices
precessing therein, assuming a constant precession frequency.
As discussed earlier in~\cite{theo1} (see also~\cite{theo2}), the
co-rotating case of the BEC problem is tantamount to the well-known
fluid case. As a result (after writing the stationary problem in the
frame rotating with the vortices), 
by defining a suitable generating function with
roots thereof representing the positions of the vortices, it is found
that this function satisfies a Hermite differential equation, indicating
that the vortices in such a rigidly rotating configuration
are sitting at the location of the well-known
roots of Hermite polynomials. 

On the other hand, the focus herein
has been on the less straightforward
case of genuinely stationary configurations consisting of oppositely charged 
vortices. For this case, we derived the appropriate generalized
differential equation, illustrating that while in the case of the
dipole, the vortex equilibrium positions are still Hermite polynomial
roots, this is not true more generally. We used the relevant differential
equation in a host of cases of different (small) numbers of positive
and negative charge vortices, to identify the corresponding equilibria
consonantly (in all cases that were examined before) to what was known
about such roots. We also used the relevant polynomial expressions to derive
``moment'' conditions about the positions of the vortices in the general
case of arbitrary vortex numbers.

While the above considerations were initially set on the line (i.e.,
collinear equilibria), subsequently, we generalized the configurations to the 
complex plane and identified relevant additions that needed to be made
to the differential equation. Once again profound similarities, but also
important differences were illustrated with respect to the classic
work of~\cite{aref11} (and references therein). 
Armed with the relevant
differential equation, and despite its complexity, 
we were able to identify one of the prototypical configurations 
of vortices in the complex plane well known in the BEC problem.
In particular, we obtained two interlaced vortex polygons, with
an equal number of vortices $n$, shifted relative to each other
by an angle of $\pi/n$.  In a similar fashion we were able to identify two nested, interlaced vortex polygons in cases with and without a vortex at the center of the configuration as well as the so-called $N+1$ vortex solution, where a single vortex of (sufficiently high -- relevant
conditions were derived--) charge $S_0$ is surrounded by vortices
of opposite charge $S$.

There are numerous directions that open up as a result of the 
analysis and calculations presented herein.

Starting from the end and the considerations in the two-dimensional
plane, it would certainly be interesting and relevant to extend
these calculations to other configurations. 
A prototypical case of especially high interest, 
as highlighted at the end of the
previous section, concerns asymmetric stationary
configurations. For these, it would be especially interesting, either
using the modified Tkachenko equation developed herein
and/or numerical approaches to be able to identify such configurations
(and, of course, to also get a sense of their potential 
stability/instabilities).

Furthermore, the exposition of~\cite{aref11} provides yet
another particularly interesting avenue of potential analysis
concerning the collinear vortices. In particular, it is found 
that a recursion
formula can be set up that provides a direct connection between
the polynomials $P_{n-1}$, $P_n$ and $P_{n+1}$ arising for different
vortex numbers. This recursion relation is associated (in the
absence of precession) to the so-called Adler-Moser polynomials~\cite{adler}
that surprisingly also arise in the study of rational solutions
of the Korteweg-de Vries equation. Identifying the relevant recursion
relation and connecting it (potentially) to particular polynomial
properties would be a separate topic worth investigating in its own
right. However, our preliminary calculations suggest that this may
not be straightforward. In such a calculation one assumes that the
two generating functions $P$ and $Q$ satisfy Eq.~(\ref{eqvor5}) and
also that $P$ and $R$ are an additional pair satisfying a similar
modified Tkachenko equation:
\begin{eqnarray}
P R'' + R P''=2 P' R' + 2 (n_+ + n_-) P R - 2 x (P R)',
\label{eqvor5a}
\end{eqnarray}
provided that $R' Q- R Q'= F(P,x)$. If an $F$ is found such that
Eqs.~(\ref{eqvor5}) and~(\ref{eqvor5a}) are concurrently satisfied, 
then $Q$, $P$ and $R$ become successive elements in 
a recursive formula that allows us to build from two of these
polynomials, the next one in the sequence of such polynomials
(i.e. $Q \sim P_{n-1}$, $P \sim P_n$ and $R \sim P_{n+1}$ and one
can then use the recursion to identify $P_{n+2}$ and so on).
In the case where Eqs.~(\ref{eqvor5}) and~(\ref{eqvor5a}) only
possess the first term of the right hand side (cf.~\cite{aref11}),
it is found that $F(P,x)=P^2$ and the Adler-Moser polynomials arise
as a result of the recursion. Here, a modification of that
calculation yields that, for example, $F(P,x)=P^2 e^{-x^2}$
provides such a solution, however the non-polynomial nature of
the exponential factor precludes its use for the purposes of
a recursive construction of higher order polynomials in this hierarchy.
Of course, this result is inconclusive as it does not immediately
preclude the presence of such polynomials, but it is perhaps suggestive
in that direction. This theme is certainly worthy of further examination,
as well.

Finally, all of the above considerations have been developed for
a constant precession frequency. Nevertheless, it has been theoretically
proposed~\cite{fetter1} and experimentally confirmed~\cite{freilich10}
that as the outer rim of the condensates is approached, the precession
frequency in fact increases. Although not accurate for all precession
radii, a reasonable approximation of this radial dependence is
$\omega= \omega_0/(1-r^2)$~\cite{navar13} 
(with the radius of precession normalized
to the size of the condensate or more concretely to the so-called
Thomas-Fermi radius). This dependence suggests the
interest and relevance of a potential generalization of the considerations
presented herein to a case of a radially dependent precession frequency.
It is then natural to expect that even in the collinear vortex case,
summation terms (involving the roots) somewhat reminiscent of
the ones arising e.g. in Eq.~(\ref{eqnew5}) will appear due to
the $1/(1-r_j^2)$ term. Nonetheless, the corresponding Tkachenko equation
may still be useful (at least in the case of small $n$'s as considered
herein) towards the identification of the vortex equilibrium positions.

Studies along these directions are currently in progress and will be 
reported in future publications.

\vspace{5mm}

{\bf Acknowledgements}: P.G.K. acknowledges support from the National Science
Foundation under grants CMMI-1000337, DMS-1312856, from FP7-People under 
grant IRSES-606096 from
the Binational (US-Israel) Science Foundation through grant 2010239, and from 
the US-AFOSR under grant FA9550-12-10332.

\end{document}